\documentclass[twocolumn,showpacs,preprintnumbers,amsmath,amssymb,floatfix,nofootinbib,superscriptaddress]{revtex4-1}
\usepackage{amsthm}
\usepackage{graphicx}
\usepackage{dcolumn,xcolor}
\usepackage{bm}
\usepackage[normalem]{ulem}
\newcommand{\be}{\begin{equation}}
\newcommand{\ee}{\end{equation}}
\newcommand{\ba}{\begin{eqnarray}}
\newcommand{\ea}{\end{eqnarray}}
\newcommand{\ban}{\begin{eqnarray*}}
\newcommand{\ean}{\end{eqnarray*}}
\newcommand{\non}{\nonumber}
\newcommand{\eq}[1]{(\ref{#1})}
\newcommand{\n}[1]{\label{#1}}
\newcommand{\hh}{\, ,\hspace{0.25cm}}
\newcommand{\hhh}{\, ,\hspace{0.5cm}}
\newcommand{\ind}[1]{\mbox{\tiny #1}}

\newcommand{\cu}{{\cal U}}

\begin{document}
 
\title{Geodesic motion around a distorted static black hole}
\author{Andrey A. Shoom}
\email{ashoom@mun.ca}
\affiliation{Department of Mathematics and Statistics, Memorial University, St. John's, Newfoundland and Labrador, A1C 5S7, Canada}
\author{Cole Walsh}
\email{cjw544@mun.ca}
\affiliation{Department of Physics and Physical Oceanography, Memorial University, St. John's, Newfoundland and Labrador, A1C 5S7, Canada}
\author{Ivan Booth}
\email{ibooth@mun.ca}
\affiliation{Department of Mathematics and Statistics, Memorial University, St. John's, Newfoundland and Labrador, A1C 5S7, Canada}

\begin{abstract}
In this paper we study geodesic motion around a distorted Schwarzschild black hole. We consider both timelike and null geodesics which are confined to the black hole's equatorial plane. Such geodesics generically exist  if the distortion field has only even interior multipole moments and so the field is symmetric with respect to the equatorial plane. We specialize to the case of distortions defined by a quadrupole Weyl moment. An analysis of the effective potential for equatorial timelike geodesics shows that finite stable orbits outside the black hole 
are possible only for $q\in(q_{\text{min}}, q_{\text{max}}]$, where $q_{\text{min}}\approx-0.0210$ and $q_{\text{max}}\approx2.7086\times10^{-4}$, while for null equatorial geodesics a finite stable orbit outside the black hole is possible only for $q\in[q_{\text{min}},0)$. Moreover, the innermost stable circular orbits (ISCOs) are closer to the distorted black hole horizon than those of an undistorted Schwarzschild black hole for $q\in(q_{\text{min}},0)$ and a null ISCO exists for $q=q_{\text{min}}$. These results shows that an external distortion of a negative and sufficiently small quadrupole moment tends to stabilize motion of massive particles and light.
\end{abstract}


\maketitle

\section{INTRODUCTION}

A study of geodesics exposes geometric properties of a curved space-time such as its isometries and hidden symmetries, which allow for integrability of geodesic equations (see, e.g., \cite{Carter,DK,FrK} and references therein). Moreover, an analysis of timelike and null geodesics, along which test particles and light propagate respectively, is useful for understanding the causal structure of a space-time (see, e.g., \cite{HE,Carter,Tipler1,Tipler2,Chandrabook,Hackmann1,Kagramanova,Diemer}) and physical processes in regions of strong gravitational field (for example, around astrophysical black holes and in the regions where gravitational waves are present). Among such physical processes are gravitational lensing (see, e.g., \cite{Perlick,FZ,Enolskii,Shohreh1,Shohreh2}), periastron shift (see, e.g., \cite{Hackmann2,Enolskii,Garcia}), the Lense-Thirring effect (see, e.g., \cite{Mashhoon1,Hackmann2}), and gravitational spinoptics (see, e.g., \cite{Mashhoon2,Mashhoon3,FrSh1,FrSh2}). 

There is an immense literature in which geodesic motion is studied and integrated analytically or numerically for many different types of a gravitational field. As a result of these studies, many interesting phenomena, such as existence of a photon sphere around a black hole (see, e.g., \cite{Chandrabook,FN,MTW,Teo,Cederbaum1,Cederbaum2,Cederbaum3}) or chaotic motion of a test particle in a nonhomogenious vacuum pp-wave solution \cite{PV1,PV2} have been found. Equatorial motion of charged particles in the vicinity of the magnetically and tidally deformed black hole represented by the Preston-Poisson metric was studied in \cite{Konoplya}.

This paper continues in this genre studying timelike and null geodesics around a static distorted black hole. Specifically we study Schwarzschild black holes distorted by static and axisymmetric gravitational fields. These space-times are members of the Weyl class of solutions.  Astrophysically such distortions would be induced by surrounding matter fields, however, these are not explicitly included in Weyl space-times. Instead the distortions are induced by a divergent asymptotic infinity. As such if we are interested in potential astrophysical effects we must restrict our attention to timelike and null geodesics in the vicinity of the black hole horizon and consider these solutions as  {\em local static distorted black holes} (see, e.g., \cite{GH,FS,AFS,Man,ManQ}).
 
  Our goal is to see how distortion affects the geodesics. In this initial study we restrict our attention to geodesics confined to the black hole's equatorial plane. Such geodesics exist generically if the distortion has reflective symmetry across the  plane. In particular we focus on the case of a quadrupole distortion. We demonstrate that stable circular and bound timelike and null geodesics exist if a quadrupole moment defining the distortion strength takes relatively small negative values. The existence of bound null geodesics is an interesting effect which was also observed in a five-dimensional vacuum solution which describes a uniform black string \cite{Gonzalez}. 

Recall that there is only one circular null geodesic orbit around a Schwarzschild black hole of mass $m$. It is located at $r=3m$ and there are no other bound null geodesics. The presence of a small quadrupole distortion slightly modifies the Schwarzschild space-time in the vicinity of the black hole horizon. As we show in this paper, such a distortion by a negative and sufficiently small quadrupole moment creates a ``corridor of light''  in the equatorial plane of the black hole,     

Besides the bound null geodesics, there are circular stable and marginally stable null and timelike geodesics around a distorted Schwarzschild black hole. Moreover, for a small negative quadrupole moment, timelike ISCOs come closer to the black hole horizon, while for a small positive quadrupole moment they move away (as compared to the ISCO of an undistorted Schwarzschild black hole). These results may have astrophysical applications as realistic black holes are surrounded by external matter and fields which create distortion effects. However, in this paper we focus only on a dynamical analysis of timelike and null geodesics.  

The paper is organized as follows. In the next section, we present the metric of a distorted Schwarzschild black hole, briefly discuss its main properties, and bring it to a computational convenient dimensionless form. Related useful expressions are presented in the Appendix. Section III contains equations for timelike and null geodesics confined to the black hole's equatorial plane. In Section IV we restrict ourselves to a quadrupole distortion and study properties of the effective potential of the timelike and null geodesic motion. Section V contains our main results, the basic properties of timelike and null geodesics. We summarize our results in the last section.

In this paper we use the following convention of units: $G=c=1$, and the sign conventions adopted in \cite{MTW}.

\section{Metric of a distorted static\\ black hole}

Let us begin with the metric representing a {\em local static distorted black hole} (see, e.g., \cite{GH,FS,AFS,Man,ManQ}), 
\ba\n{II.1}
ds^{2}&=&-\left(\frac{x-1}{x+1}\right)e^{2U}dt^{2}+m^{2}(x+1)^{2}(1-y^{2})e^{-2U}d\phi^{2}\non\\
&+&m^{2}(x+1)^{2}e^{2(V-U)}\left(\frac{dx^{2}}{x^{2}-1}+\frac{dy^{2}}{1-y^{2}}\right)\,.
\ea
Here $t\in(-\infty, +\infty)$ is the time coordinate, $x\in(1, +\infty)$ and $y\in[-1, 1]$ are the prolate spheroidal coordinates, and $\phi\in[0, 2\pi)$ is the azimuthal angular coordinate. The metric functions $U(x,y)$ and $V(x,y)$ represent a distortion. They solve the vacuum Einstein equations \eq{A1}--\eq{A3} and are given in terms of the Legendre polynomials of the first kind,
\ba
U&=&\sum_{n\geq0}a_{n}R^{n}P_{n}\,,\n{II.2a}\\
V&=&\sum_{n,k\geq1}\frac{nka_{n}a_{k}}{(n+k)}R^{n+k}(P_{n}P_{k}-P_{n-1}P_{k-1})\,\non\\
&+&\sum_{n\geq1}a_{n}\sum_{l=0}^{n-1}[(-1)^{n-l+1}(x+y)-x+y]R^{l}P_{l}\,.\n{II.2b}\\
P_{n}&\equiv&P_{n}(xy/R)\hh R=\sqrt{x^{2}+y^{2}-1}\,.\n{II.3}
\ea
The Legendre polynomials have the following properties useful for our calculations: 
\ba
P_n(-x)&=&(-1)^nP_n(x)\hh P_n(1)=1\,,\n{II.4a}\\
\frac{dP_{n}(x)}{dx}&=&\frac{nP_{n-1}(x)-nxP_{n}(x)}{1-x^{2}}\,,\n{II.4b}\\
P_{2k}(0)&=&(-1)^{k}\frac{(2k-1)!!}{(2k)!!}\hh P_{2k+1}(0)=0\,,\n{II.4c}\\
k&=&0,1,2,3,...\,,\non\\
(2k-1)!!&=&1\cdot3\cdot5\cdot...\cdot(2k-1)\hh (-1)!!=1\,,\non\\
(2k)!!&=&2\cdot4\cdot6\cdot...\cdot(2k)\hh 0!!=1\,.\non
\ea
The black hole's horizon is located at $x=1$ and the space-time singularity is located at $x=-1$. The metric functions $U$ and $V$ calculated on the horizon are the following \cite{AAS}:
\be\n{II.4d}
U(1,y)=\sum_{n\geq0}a_{n}y^{n}\hh V(1,y)=2(U(1,y)-u_{0})\,,
\ee
where
\be\n{II.4e}
u_{0}=\sum_{n\geq0}a_{n}\,.
\ee

The multipole moments $a_{n}$ define a distortion\footnote{These multipole moments sometimes are called the Weyl multipole moments. A relation of the Weyl multipole moments to their relativistic analogues was discussed in \cite{SuenI} for the Schwarzschild black hole distorted by an external field. The general formalism, which includes both the Thorne \cite{Thorne} and the Geroch-Hansen \cite{Ger1,Ger2,Han,Que} relativistic multipole moments is presented in \cite{SuenII}. A relation between the Thorne and the Geroch-Hansen relativistic multipole moments is given in \cite{Beig,Gursel}.} which is due to an external static gravitational field (see, e.g., \cite{Man,ManQ}). According to the terminology used in Newtonian gravitational theory and electromagnetism, the coefficients in the multipole expansion of the distortion gravitational field are called {\em interior multipole moments}.\footnote{Note that even though $U$ satisfies the Laplace equation, it is a relativistic field. In order to construct the corresponding Newtonian analogue of the field, one has to take the nonrelativistic limit $\lim_{c^{2}\to\infty}c^{2}U(x,y,c^{2})$, where $c$ is the speed of light (see, e.g., \cite{Eh,Qu}).} The distortion fields $U$ and $V$ defined by the interior multipole moments are regular and smooth at the black hole horizons. {\em Exterior multipole moments} describe deformations of the source \cite{Man,ManQ}. They are given in terms of the Legendre polynomials of the second kind (see, e.g., \cite{Dor,ManQ}). According to the Schwarzschild black hole uniqueness theorem \cite{Israel}, the Schwarzschild black hole is the only static, asymptotically flat, vacuum black hole with a regular horizon. Thus, such deformations make the black hole horizon singular (see, e.g., \cite{Dor}). 

The first term in the expansion of the distortion field $U$ is the monopole and in our case it represents a background distortion defined by a monopole moment $a_{0}$. The next term is the dipole defined by a dipole moment $a_{1}$, which according to the black hole equilibrium condition (see Eq.\eq{II.7} below), is related to the higher order multipole moments.  The next term is the quadrupole, which is defined by a quadrupole moment $a_{2}$. Here we shall consider only the subleading terms in the multipole expansion of the distortion field.

To have the horizon free of conical singularities on the symmetry axis $y=\pm1$ the multipole moments have to satisfy the following condition:
\be\n{II.7}
\sum_{n\geq0}a_{2n+1}=0\,.
\ee
This condition is sometimes called the black hole equilibrium condition \cite{Chandrabook}.

If we consider that the distortion field is generated by some material sources which satisfy the strong energy condition, then we necessarily have (see, e.g. \cite{GH}), 
\be\n{II.8a}
U\leq0\,,
\ee
which implies 
\be\n{II.8b}
u_{0}\leq0\hh u_{0}=\sum_{n\geq0}a_{2n}\,.
\ee
The expressions above define a {\em local static distorted black hole}. 

Using the coordinate transformations 
\be\n{II.9}
x=\frac{r}{m}-1\hhh y=\cos\theta\,,
\ee
and removing the distortion by making all the multipole moments $a_{n}$ vanish we derive the Schwarzschild metric (see, e.g., \cite{MTW,FN}),
\ba\n{II.10}
ds^2&=&-\left(1-\frac{2m}{r}\right)dt^2+\left(1-\frac{2m}{r}\right)^{-1}dr^2+r^{2}d\omega^{2}\,,\non\\
d\omega^{2}&=&d\theta^2+\sin^2\theta d\phi^2\,.
\ea

In what follows, it is convenient to present the metric \eq{II.1} in a dimensionless form
\ba\n{II.5}
dS^{2}&=&-\left(\frac{x-1}{x+1}\right)e^{2\cu}dT^{2}+(x+1)^{2}(1-y^{2})e^{-2\cu}d\phi^{2}\non\\
&+&(x+1)^{2}e^{2(V-\cu)}\left(\frac{dx^{2}}{x^{2}-1}+\frac{dy^{2}}{1-y^{2}}\right)\,,
\ea
where
\ba\n{II.6}
ds^{2}&=&\Omega^{2}dS^{2}\hh\Omega^{2}=m^{2}e^{-2u_{0}}\,,\non\\
t&=&me^{-2u_{0}}T\hh \cu=U-u_{0}\,.
\ea

\section{Geodesic equations}

In this section we construct equations for timelike and null geodesics in the vicinity of a distorted static black hole. We restrict ourselves to the geodesics lying in the equatorial plane of the distorted black hole. As we illustrate below, such geodesics exist only for certain types of distortion. 

A geodesic equation is defined as follows:
\be\n{III.1}
\ddot{x}^{\alpha}+\Gamma^{\alpha}_{\,\,\,\beta\gamma}\dot{x}^{\beta}\dot{x}^{\gamma}=0\,.
\ee
Here the overdot stands for the derivative with respect to a proper time $\tau$, for timelike geodesics, or with respect to an affine parameter $\lambda$, for null geodesics. For the space-time \eq{II.5} the Christoffel symbols are the following:
\ba\n{III.2}
\Gamma^{x}_{TT}&=&\frac {( x-1)}{( x+1) ^{3}}\left[1+ ({x}^{2}-1)\,\cu_{,x}\right]e^{4\cu-2V}\,,\non\\
\Gamma^{x}_{xx}&=&-\frac{1}{x^{2}-1}+V_{,x}-\cu_{,x}\,,\non\\
\Gamma^{x}_{xy}&=&V_{,y}-\cu_{,y}\,,\non\\
\Gamma^{x}_{yy}&=&-\frac {( x-1)}{(1-y^{2})}\left[1+(x+1)(V_{,x}-\cu_{,x})\right]\,,\non\\
\Gamma^{x}_{\phi\phi}&=&-(x-1)(1-y^{2})\left[1-(x+1)\,\cu_{,x}\right]e^{-2V}\,,\\
\Gamma^{y}_{TT}&=&\frac{( x-1)}{(x+1) ^{3}}(1-{y}^{2})\,\cu_{,y}e^{4\cu-2V}\,,\non\\ 
\Gamma^{y}_{xx}&=&\frac{(1-y^{2})}{(x^{2}-1)}\left(\cu_{,y} -V_{,y}\right)\,,\non\\
\Gamma^{y}_{xy}&=&\frac{1}{x+1}+V_{,x}-\cu_{,x}\,,\non\\
\Gamma^{y}_{yy}&=&\frac{y}{1-y^{2}}+V_{,y}-\cu_{,y}\,,\non\\
\Gamma^{y}_{\phi\phi}&=&(1-y^{2})\left[y+(1-y^{2})\,\cu_{,y}\right]e^{-2V}\,.\non
\ea

The space-time \eq{II.5} has Killing vectors $\xi^{\alpha}_{(T)}=\delta^{\alpha}_{T}$ and $\xi^{\alpha}_{(\phi)}=\delta^{\alpha}_{\phi}$. Accordingly, there are the following quantities conserved along a geodesic:
\ba
{\cal E}&\equiv&-\xi^{\alpha}_{(T)}u_{\alpha}=\left(\frac{x-1}{x+1}\right)e^{2\cu}\,\dot{T}\,,\n{III.3a}\\
{\cal L}_{\phi}&\equiv&\xi^{\alpha}_{(\phi)}u_{\alpha}=(x+1)^{2}(1-y^{2})e^{-2\cu}\dot{\phi}\,.\n{III.3b}
\ea
For a massive particle the quantity ${\cal L}_{\phi}$ represents the particle's dimensionless azimuthal angular momentum. The interpretation of ${\cal E}$ is more complicated. For an asymptotically flat space-time 
with $\xi^\alpha_{(T)}$ normalized at asymptotic infinity it would represent the particle's per unit mass as measured by a (static)  observer at infinity moving with four-velocity $\xi^\alpha_{(T)}$. 
However our space-time is not asymptotically flat and so $\xi^\alpha_{(T)}$ does not have a natural scaling. 

While the exterior region of the space-time can be surgically modified to include
an asymptotically flat region this is not sufficient to uniquely determine the scaling: the surgery itself is not unique. Nevertheless for any given scaling ${\cal E}$ is conserved and this is enough
to proceed with the calculations. Here we choose to work with the (convenient) naive scaling inherited from the coordinate system.

With the expressions \eq{III.3a}--\eq{III.3b} the dynamical system \eq{III.1} reduces to a two-dimensional one, confined to a $(x,y)-$plane,
\ba
\ddot{x}&+&\Gamma^{x}_{xx}\dot{x}^{2}+2\Gamma^{x}_{xy}\dot{x}\dot{y}+\Gamma^{x}_{yy}\dot{y}^{2}\n{III.4a}\\
&+&{\cal E}^{2}\left(\frac{x+1}{x-1}\right)^{2}e^{-4\cu}\Gamma^{x}_{TT}+\frac{{\cal L}^{2}_{\phi}e^{4\cu}}{(x+1)^{4}(1-y^{2})^{2}}\Gamma^{x}_{\phi\phi}=0\,,\non\\
\ddot{y}&+&\Gamma^{y}_{xx}\dot{x}^{2}+2\Gamma^{y}_{xy}\dot{x}\dot{y}+\Gamma^{y}_{yy}\dot{y}^{2}\n{III.4b}\\
&+&{\cal E}^{2}\left(\frac{x+1}{x-1}\right)^{2}e^{-4\cu}\Gamma^{y}_{TT}+\frac{{\cal L}^{2}_{\phi}e^{4\cu}}{(x+1)^{4}(1-y^{2})^{2}}\Gamma^{y}_{\phi\phi}=0\,.\non
\ea
There is an additional constraint $u^{\alpha}u_{\alpha}=\varepsilon$, where $u^{\alpha}=\dot{x}^{\alpha}$ is 4-velocity, $\varepsilon=-1$ for timelike, and $\varepsilon=0$ for null geodesics. The constraint takes the following form:
\ba\n{III.5}
&&(x+1)^{2}e^{2(V-\cu)}\left(\frac{\dot{x}^{2}}{x^{2}-1}+\frac{\dot{y}^{2}}{1-y^{2}}\right)\\
&&-{\cal E}^{2}\left(\frac{x+1}{x-1}\right)e^{-2\cu}+\frac{{\cal L}^{2}_{\phi}e^{2\cu}}{(x+1)^{2}(1-y^{2})}=\varepsilon\,.\non
\ea

Let us now consider geodesics confined to the black hole's equatorial plane $(y=0)$. In order to have $y=const$, one must have $\dot{y}=\ddot{y}=0$ along such a geodesic. According to the expressions \eq{III.2}, \eq{III.4b}, \eq{III.5}, and \eq{A3}, this condition boils down to the condition
\be\n{III.6}
\cu_{,y}|_{y=0}=0\,.
\ee
Using the expressions \eq{II.2a}, \eq{II.3}, \eq{II.4b}, and \eq{II.4c} we can present this condition in an explicit form
\be\n{III.7}
\cu_{,y}=\sum_{n\geq0}a_{2n+1}x(1-x^{2})^{n}\frac{(2n+1)!!}{(2n)!!}=0\,.
\ee
This condition is satisfied generically if a distortion does not contain odd multipole moments. In other words, there are geodesics confined to the equatorial plane if
\be\n{III.8}
a_{2n+1}=0\hh n=0,1,2,...\,.
\ee
Note that this condition is stronger than the condition \eq{II.7}. Physically it corresponds to the intuitive requirement that the space-time have reflective symmetry across the equatorial plane. 
Without such symmetry one would generically expect that particles would be pulled out of that plane. 

In what follows, we shall study the equatorial geodesics. The distortion fields satisfying the condition \eq{III.8} take the following form on the equatorial plane:
\ba
\bar{\cu}&\equiv&\cu|_{y=0}=\sum_{n\geq0}a_{2n}(1-x^{2})^{n}\frac{(2n-1)!!}{(2n)!!}-u_{0}\,,\n{III.9a}\\
\bar{V}&\equiv&V|_{y=0}=-2\sum_{n\geq1}a_{2n}x\sum_{l=0}^{n-1}(1-x^{2})^{l}\frac{(2l-1)!!}{(2l)!!}\,\n{III.9b}\\
&+&2\sum_{n,k\geq1}\frac{nka_{2n}a_{2k}}{(n+k)}(1-x^{2})^{n+k}\frac{(2n-1)!!(2k-1)!!}{(2n)!!(2k)!!}\,,\non
\ea  
and the geodesic equation \eq{III.4a} reads
\ba
&&\ddot{x}-\frac{\dot{x}^{2}}{x^{2}-1}+(\bar{V}_{,x}-\bar{\cu}_{,x})\dot{x}^{2}+\frac{{\cal E}^{2}e^{-2\bar{V}}}{x^{2}-1}[1+(x^{2}-1)\,\bar{\cu}_{,x}]\non\\
&&-\frac{{\cal L}^{2}(x-1)e^{4\bar{\cu}-2\bar{V}}}{(x+1)^{4}}[1-(x+1)\,\bar{\cu}_{,x}]=0\,.\n{III.10}
\ea
Here and in what follows, ${\cal L}={\cal L}_{\phi}(y=0)$ is the dimensionless total angular momentum. The constraint \eq{III.5} projected on the equatorial plane is an integral of the geodesic equation. We present the constraint in the following archetypal form:
\ba
e^{2\bar{V}}\dot{x}^{2}&=&{\cal E}^{2}-U_{\text{eff}}\,,\n{III.11a}\\
U_{\text{eff}}&=&\left(\frac{x-1}{x+1}\right)e^{2\bar{\cu}}\left[-\varepsilon+\frac{{\cal L}^{2}\,e^{2\bar{\cu}}}{(x+1)^{2}}\right]\,.\n{III.11b}
\ea
Here $U_{\text{eff}}$ is the effective potential of the dynamical motion \eq{III.10}. We shall study its properties in the following section.

\section{Effective potential}

Let us now study properties of the effective potential \eq{III.11b}. We shall consider the case of timelike geodesics, $\varepsilon=-1$, and null geodesics, $\varepsilon=0$, separately. 

\subsection{Effective potential for timelike geodesics}

\begin{figure}[htb]
\begin{center}
\hspace{0cm}
\includegraphics[width=7.0cm]{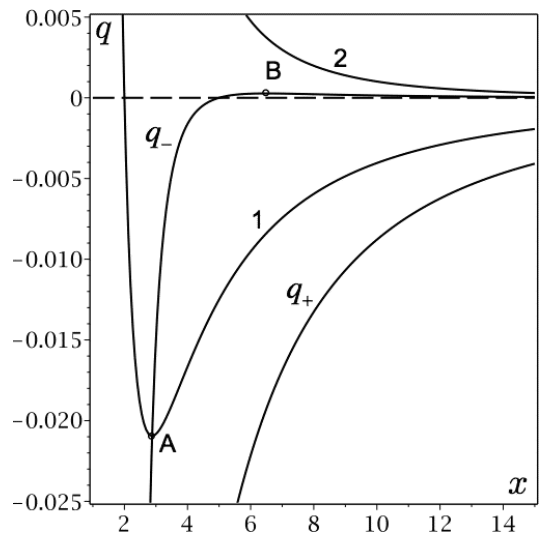}
\caption{The region enclosed between curve 1 [see Eq.\eq{IV.6}] and curve 2 [see Eq.\eq{IV.7}] defines the values of $q$ and $x$ for which extrema of $U_{\text{eff}}$ exist. Curve 1 has a minimum (point A) at $x=x_{\text{min}}=1+2\cos(\pi/9)\approx2.8794$ where $q=q_{\text{min}}\approx-0.0210$. Curves $q_{\pm}$ define extrema of the function $f$. Curve $q_{-}$ intersects curve 1 at the point $A$, it vanishes at $x=5$, and has a maximum (point B) at $x_{\text{max}}\approx6.5018$ where $q_{-}=q_{\text{max}}\approx2.7086\times10^{-4}$.} \label{f1}
\end{center}
\end{figure}

The effective potential for timelike geodesics reads
\be\n{IV.1}
U_{\text{eff}}=\left(\frac{x-1}{x+1}\right)e^{2\bar{\cu}}\left[1+\frac{{\cal L}^{2}\,e^{2\bar{\cu}}}{(x+1)^{2}}\right]\,.
\ee
In the absence of distortion this potential coincides with the effective potential for a massive test particle moving in an equatorial plane of a Schwarzschild black hole (see, e.g., p.639 \cite{MTW}). Here we shall consider a quadrupole distortion defined by a quadrupole moment $a_{2}=q$ [cf. \eq{III.9a} and \eq{III.9b}],
\ba
\bar{\cu}&=&-\frac{q}{2}(x^{2}+1)\,,\n{IV.2}\\
\bar{V}&=&-2qx+\frac{q^{2}}{4}(x^{2}-1)^{2}\,.\n{IV.2a}
\ea
Let us study properties of the corresponding effective potential in the external to the black hole region $x\in(1,+\infty)$. We have
\be\n{IV.3}
\lim_{x\to1^{+}}U_{\text{eff}}=0^{+}\hh \lim_{x\to+\infty}U_{\text{eff}}=\begin{cases}
+\infty\hh q<0\\
1^{-}\hh q=0\,\,.\\
0^{+}\hh q>0\
\end{cases}
\ee
The effective potential is positive and continuous in the region. By continuity of $U_{\text{eff}}$ we can infer that for $q\leq0$ there is an even number of extrema, if any, while for $q>0$ the number of extrema, if any, is odd. The next step is to find extrema of $U_{\text{eff}}$. An equation for the extrema $U_{\text{eff},x}=0$ is equivalent to the equation 
\be
f={\cal L}^{2}e^{-q(x^{2}+1)}\,,\n{IV.4a}
\ee
where
\be
f=\frac{(x+1)^{2}[1-qx(x^{2}-1)]}{[x-2+2qx(x^{2}-1)]}\,.\n{IV.4b}
\ee
Thus, to find extrema of the effective potential we have to study properties of the function $f$. We have
\be\n{IV.5}
\lim_{x\to1^{+}}f=-4^{-}\hh \lim_{x\to+\infty}f=
\begin{cases}
-\infty\hh q\ne0\\
+\infty\hh q=0\,.
\end{cases}
\ee
For $q=0$ the function $f$ diverges at $x=2$ and for $q\ne 0$ it diverges along the curve (see curve 1 in Fig.~\ref{f1})
\be\n{IV.6}
q=\frac{(2-x)}{2x(x^{2}-1)}
\ee
and vanishes along the curve (see curve 2 in Fig.~\ref{f1})
\be\n{IV.7}
q=\frac{1}{x(x^{2}-1)}\,.
\ee
There is one trivial solution to Eq. \eq{IV.4a} defined by ${\cal L}=0$ and Eq. \eq{IV.7}. This solution corresponds to a particle at rest, $(x,y,\phi)=const.$, with respect to the black hole. Such a solution doesn't exist for an undistorted Schwarzschild black hole, but it is possible in the case of a distorted Schwarzschild black hole where the black hole's gravitational attraction is balanced by the external distortion. 

The region enclosed between curves 1 and 2 defines the values of $q$ and $x$ for which, given appropriate values of $|{\cal L}|$, extrema of $U_{\text{eff}}$ exist. Accordingly, the effective potential has extrema if $q>q_{\text{min}}$. Let us now define the number of the extrema. To do it, we shall study extrema of the function $f$. The equation for extrema $f_{,x}=0$ is equivalent to the equation
\be\n{IV.8}
4x^{2}(x^{2}-1)^{2}q^{2}+2x(x+1)(2x^{2}-5x+4)q-x+5=0\,.
\ee
Solving this equation for $q$ we derive
\ba
q_{\pm}&=&-\frac{2x^{2}-5x+4\pm\sqrt{D}}{4x(x-1)(x^{2}-1)}\,,\n{IV.9a}\\
D&=&4x^{4}-16x^{3}+13x^{2}+4x-4\,.
\ea
The branches $q_{\pm}$ are real if $D\geq0$. This condition holds in the regions $x\in(1,x_{1})\cup(x_{2},+\infty)$, where $x_{1}\approx1.2255$, $x_{2}\approx2.7194$. The branch $q_{-}$ passes through the region enclosed by the curves 1 and 2 and intersects curve 1 at the point $A$ (see Fig.~\ref{f1}). Thus, it corresponds to the extrema of the function $f$ where $f>0$. In that region this branch has also zero value at $x=5$ and an extremum (maximum) at $x=x_{\text{max}}\approx6.5018$ where $q_{-}=q_{\text{max}}\approx2.7086\times10^{-4}$. 

\begin{figure*}[htb]
\begin{center}
\hspace{0cm}
\ba
&&\hspace{0cm}\includegraphics[width=6.0cm]{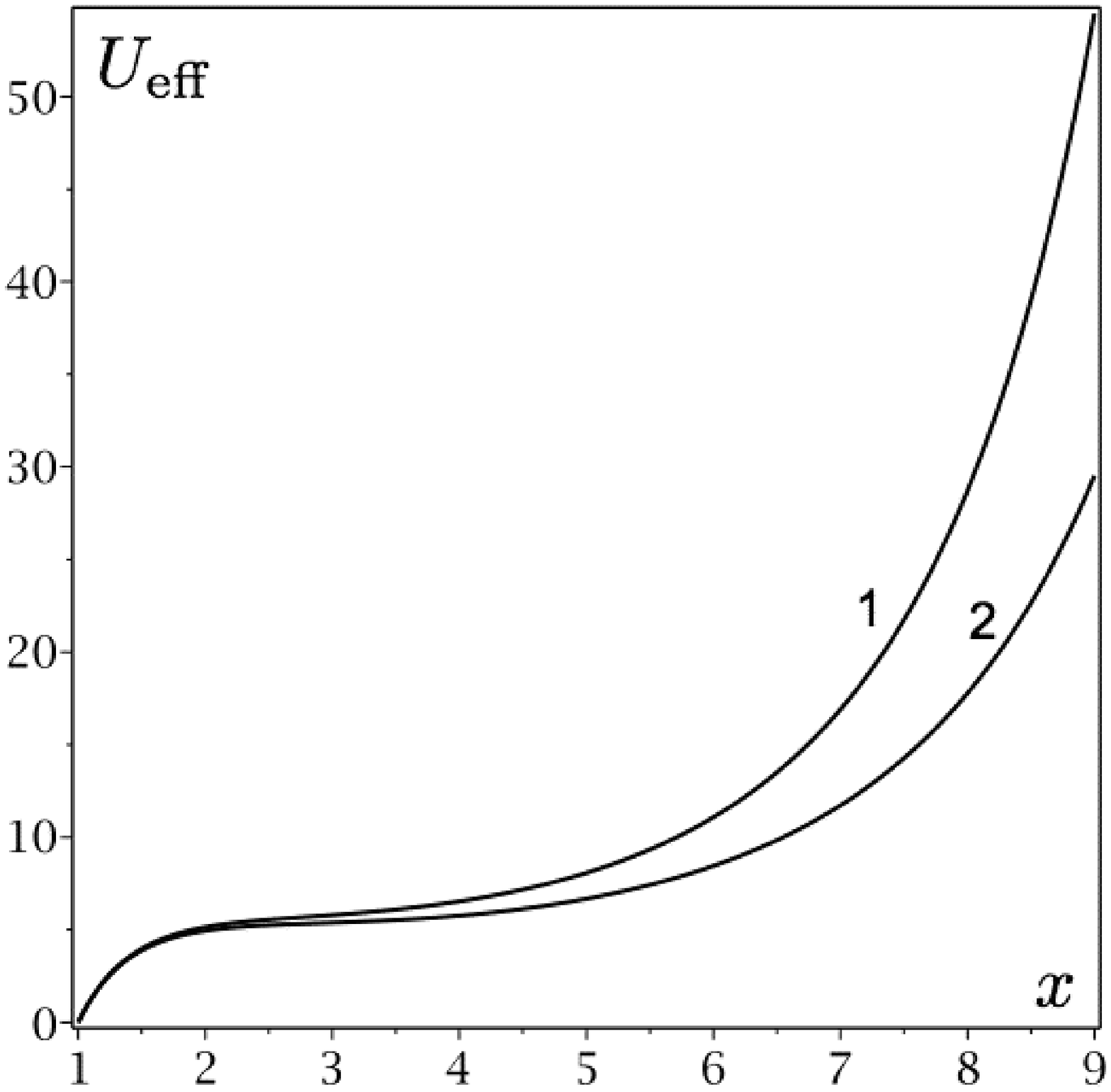}
\hspace{1.5cm}\includegraphics[width=6.0cm]{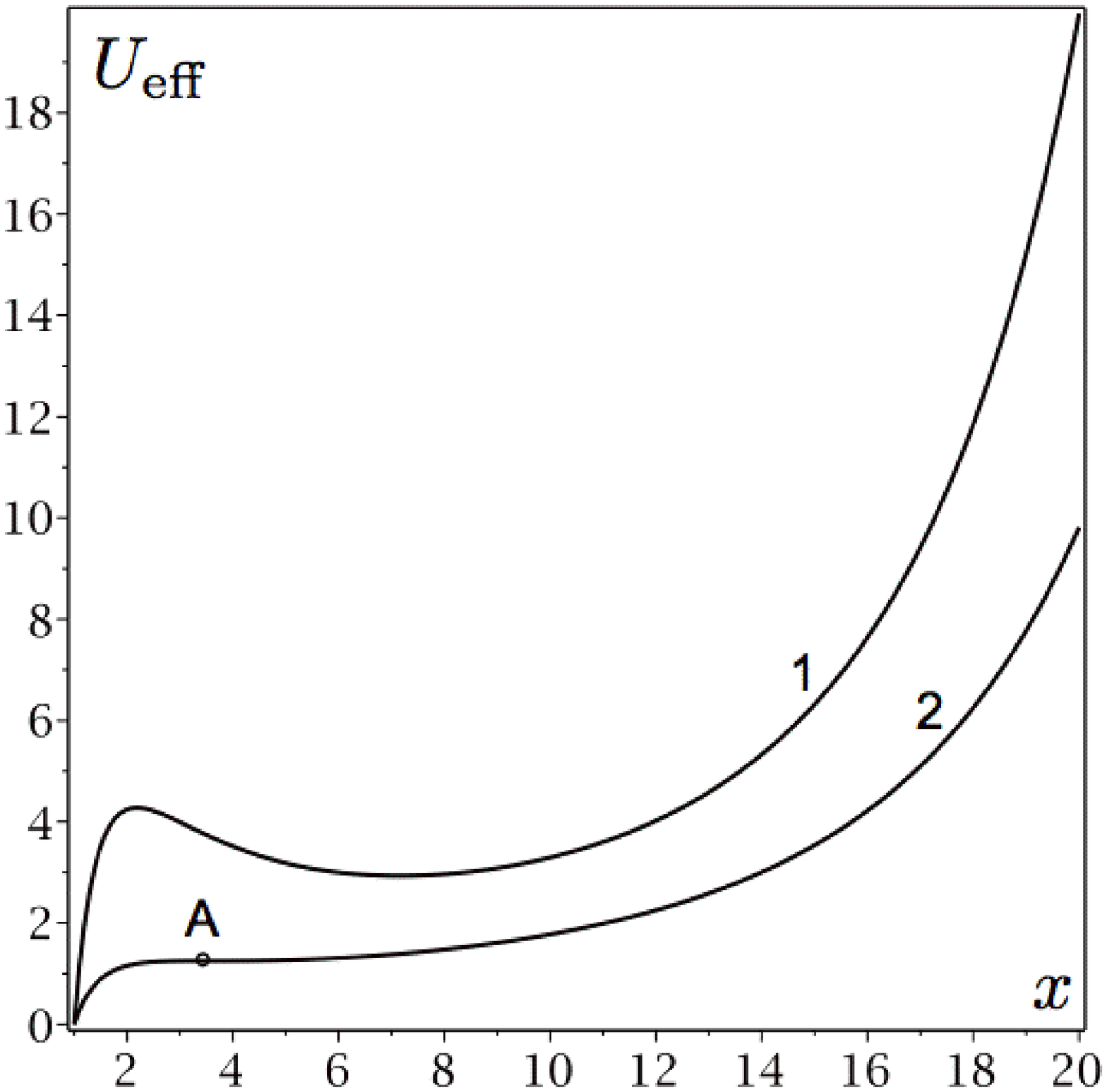}\non\\
&&\hspace{3cm}({\bf a})\hspace{7.1cm}({\bf b})\non\\
&&\hspace{0cm}\includegraphics[width=6.0cm]{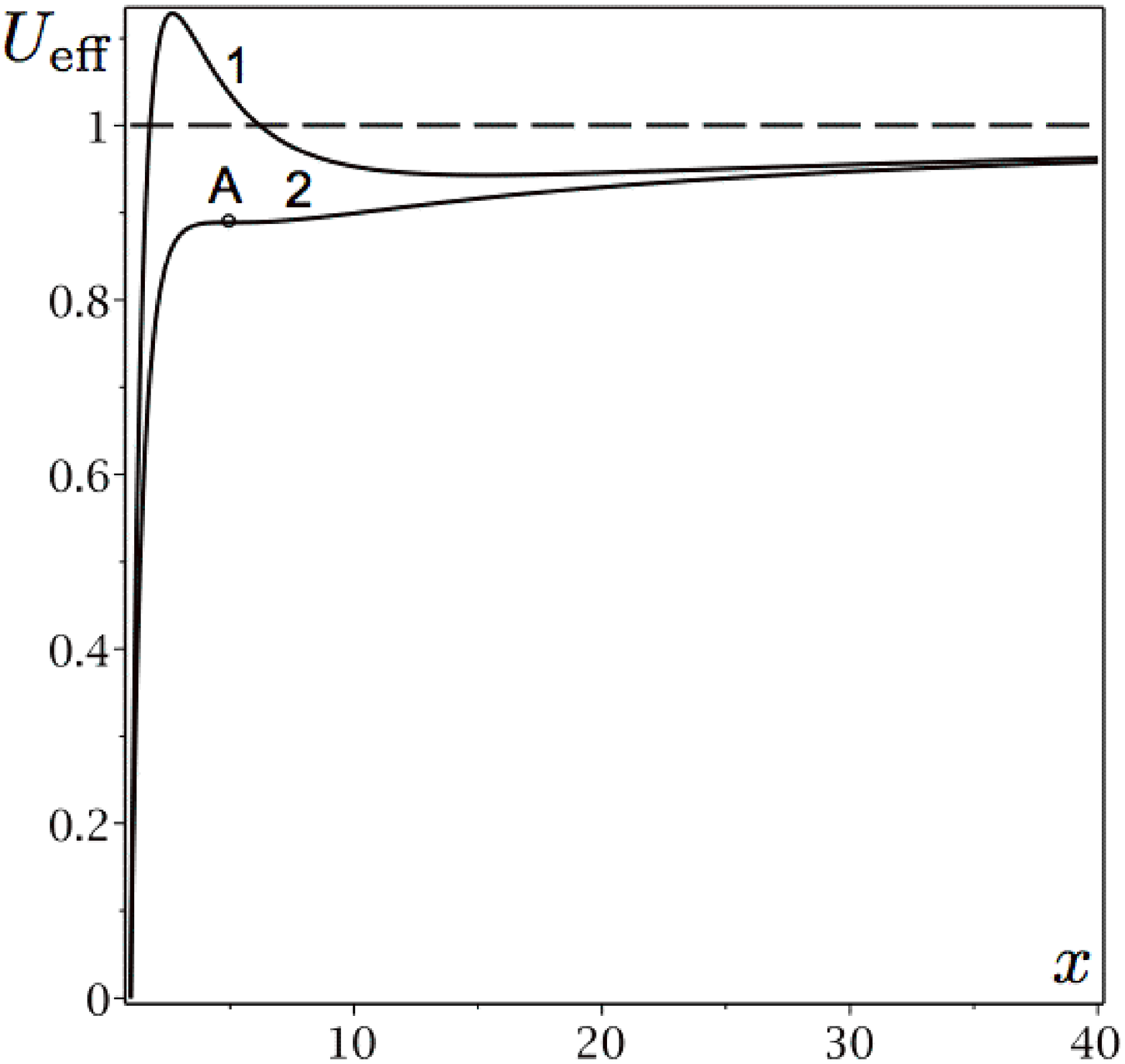}
\hspace{1.5cm}\includegraphics[width=6.0cm]{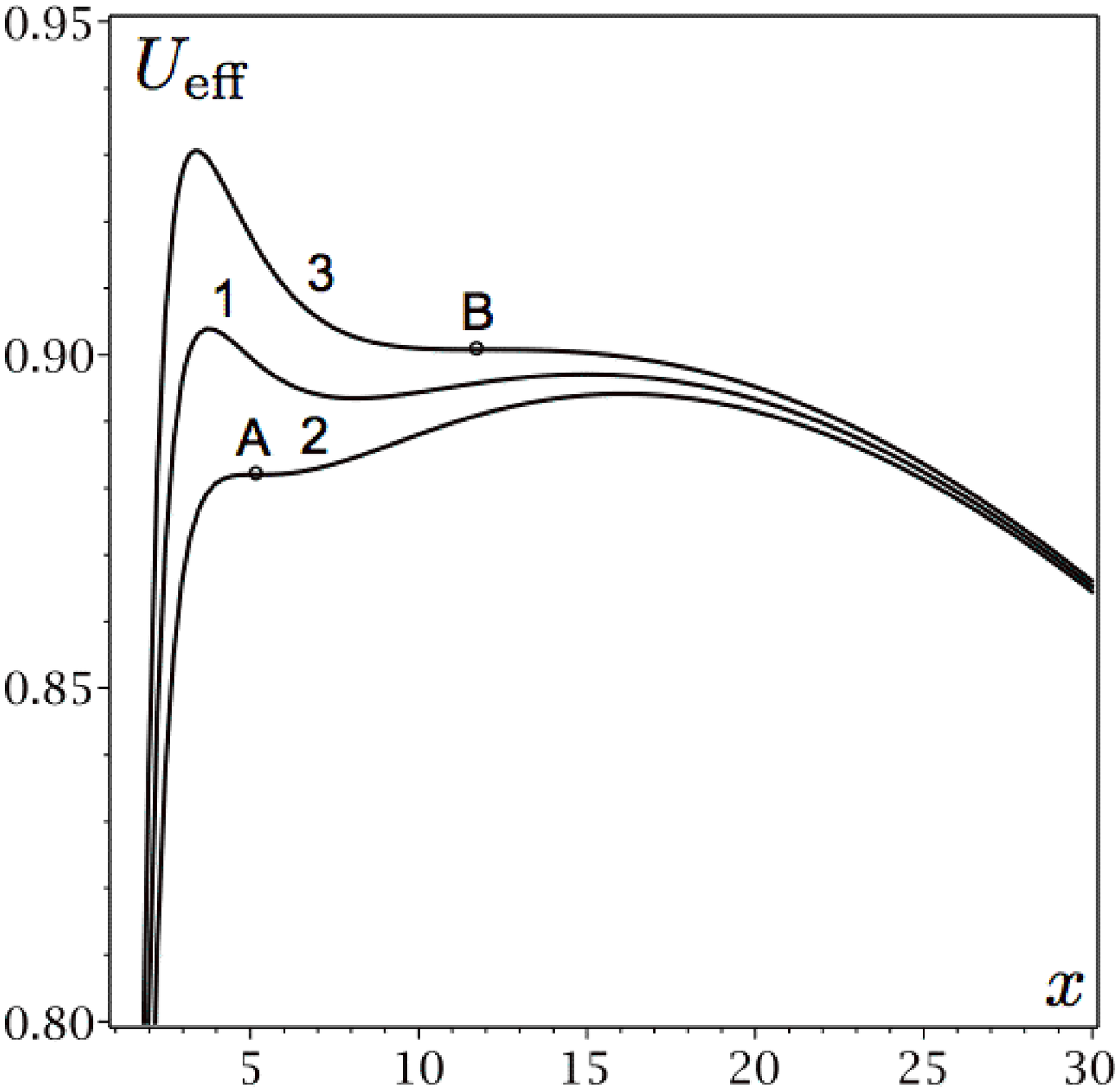}\non\\
&&\hspace{3cm}({\bf c})\hspace{7.1cm}({\bf d})\non\\
&&\hspace{0cm}\includegraphics[width=6.0cm]{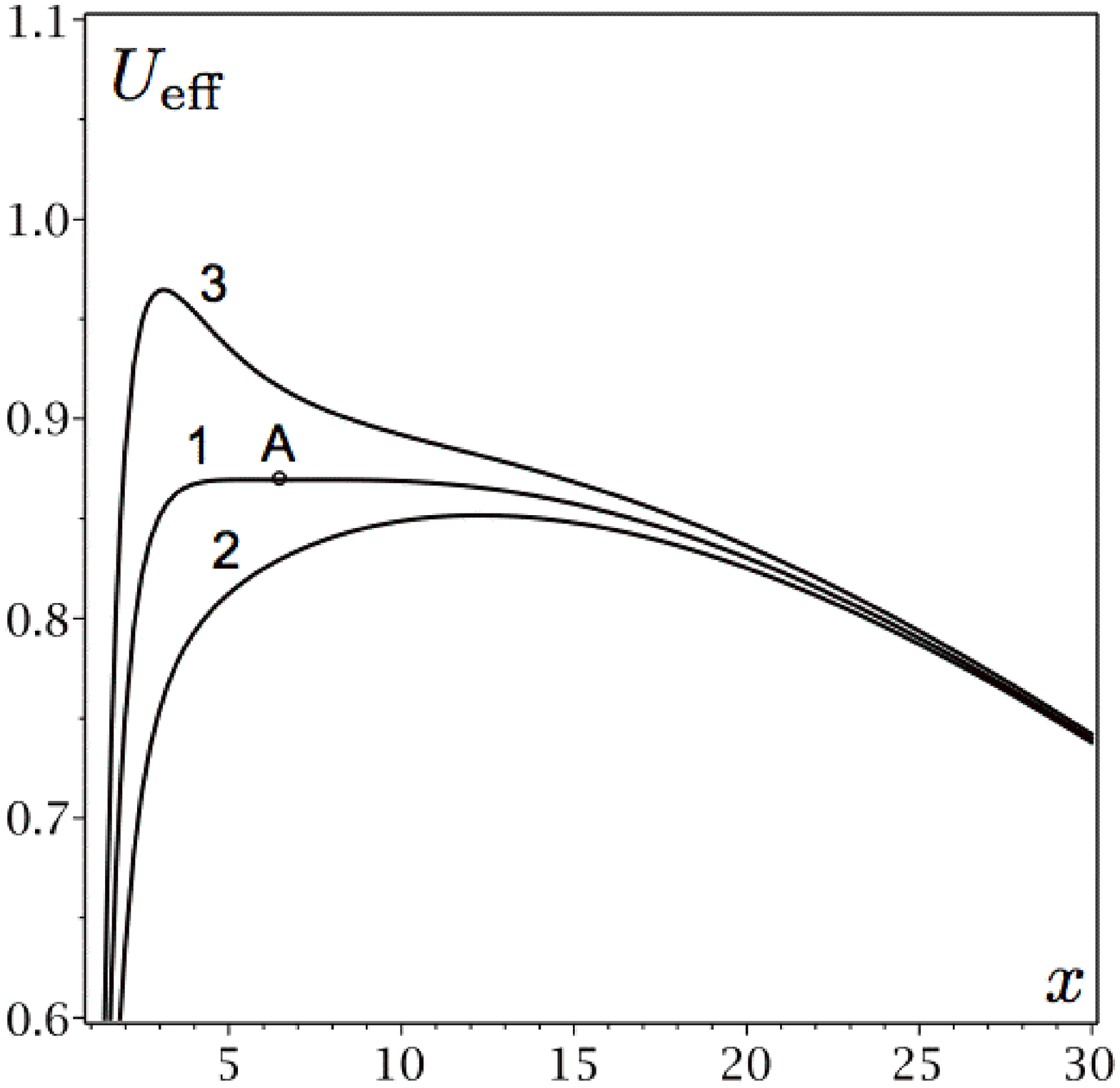}
\hspace{1.5cm}\includegraphics[width=6.0cm]{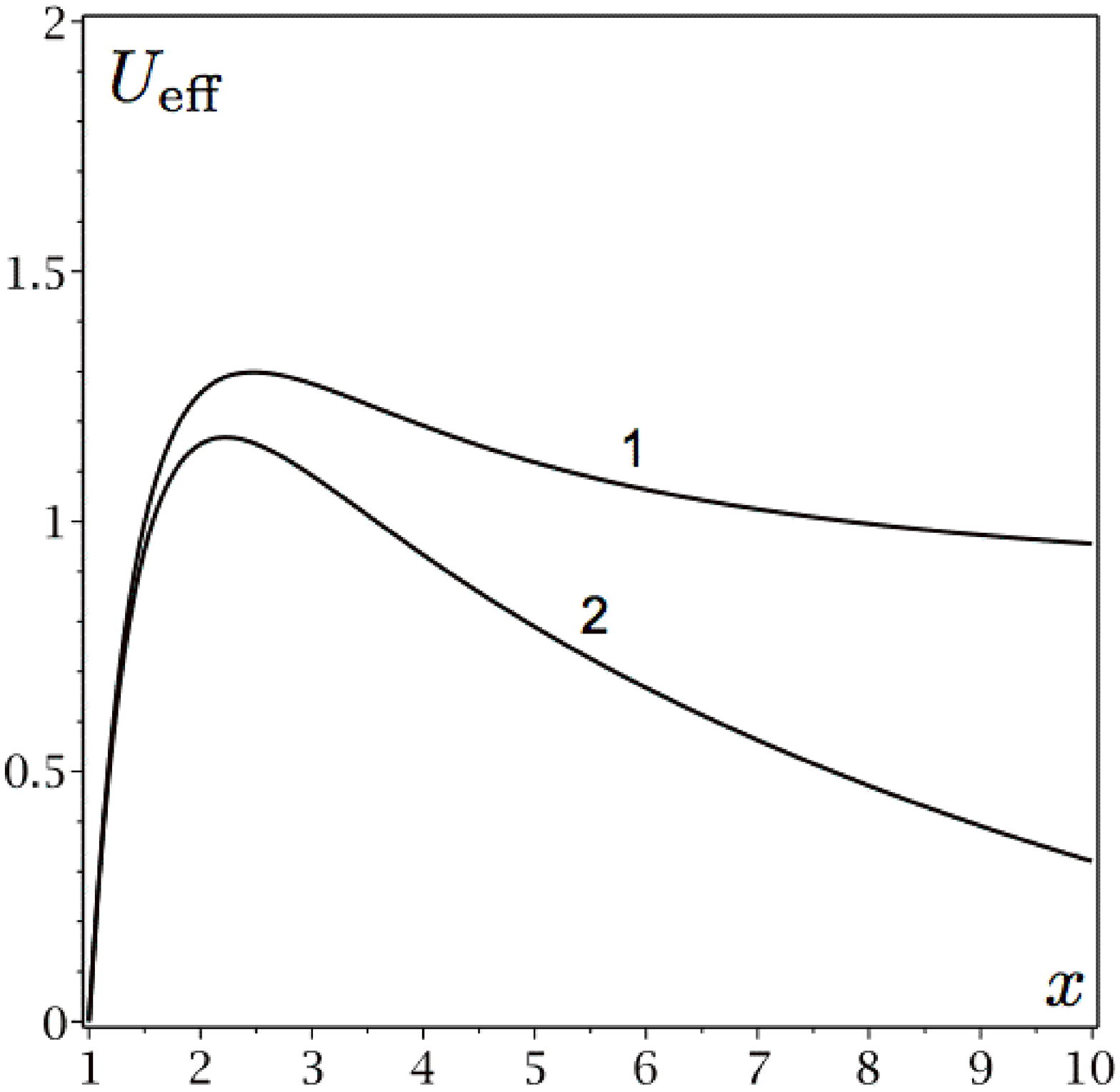}\non\\
&&\hspace{3cm}({\bf e})\hspace{7.1cm}({\bf f})\non
\ea
\caption{The effective potential for timelike geodesics. Plot (${\bf a}$): $q\leq q_{\text{min}}$, $|{\cal L}|=10$. Curve 1: $q=-0.0250$; Curve 2: $q=q_{\text{min}}$. Plot (${\bf b}$): $q\in(q_{\text{min}},0)$, $q\approx-0.0052$. Curve 1: $|{\cal L}|=10$; Curve 2: $|{\cal L}|\approx4.5625$. The inflection point $A$ is at $x=3.46$. Plot (${\bf c}$): $q=0$. Curve 1: $|{\cal L}|=1+2\sqrt{3}$; Curve 2: $|{\cal L}|=2\sqrt{3}$. The inflection point $A$ is at $x=5$. Plot (${\bf d}$): $q\in(0,q_{\text{max}})$, $q=0.0001$. Curve 1: $|{\cal L}|\approx3.5691$; Curve 2: $|{\cal L}|\approx3.4327$. The inflection point $A$ is at $x\approx5.2092$;  Curve 3: $|{\cal L}|\approx3.7055$. The inflection point $B$ is at $x\approx11.7595$. Plot (${\bf e}$): $q=q_{\text{max}}$. Curve 1: $|{\cal L}|\approx3.3708$. The inflection point $A$ is at $x=x_{\text{max}}\approx 6.5018$; Curve 2: $|{\cal L}|\approx2.8708$; Curve 3: $|{\cal L}|\approx3.8708$. Plot (${\bf f}$): $q\geq q_{\text{max}}$, $|{\cal L}|=5$. Curve 1: $q=q_{\text{max}}$; Curve 2: $q=0.01$.} \label{f2}
\end{center}
\end{figure*}

Using Fig.~\ref{f1} we can deduce properties of the function $f$ and, consequently, properties of the effective potential. They are as follows: 
\begin{description}
\item[$q<q_{\text{min}}$] The function $f$ is negative, thus the effective potential doesn't have extrema for any value of $|{\cal L}|$. According to \eq{IV.3}, it grows monotonically in the region $x\in(1,+\infty)$ [see Fig.~\ref{f2}, plot (${\bf a}$)].
\item[$q=q_{\text{min}}$] The function $f$ diverges at $x=x_{\text{min}}$. It has one extremum (maximum) in the region $x\in(x_{\text{min}},+\infty)$. The function $f$ is negative in the region $x\in(1,+\infty)$. Accordingly, Eq. \eq{IV.4a} has no real solutions. Thus, the effective potential has no extrema for any value of $|{\cal L}|$. Its behaviour is the same as in the previous case [see Fig.~\ref{f2}, plot (${\bf a}$)].
\item[$q\in(q_{\text{min}},0)$] The function $f$ diverges at two points, $x=x_{I}$ and $x=x_{II}$ given by roots of Eq. \eq{IV.6}. It has one extremum (minimum) at $x\in(x_{I},x_{II})$, where it is positive, and one extremum (maximum) at $x\in(x_{II},+\infty)$, where it is negative. Thus, Eq. \eq{IV.4a} has at most two real roots for appropriate values of $|{\cal L}|>|{\cal L}_{\text{min}}|$, where $|{\cal L}_{\text{min}}|$ depends on the value of $q$. These roots merge for $|{\cal L}|=|{\cal L}_{\text{min}}|$. Accordingly, the effective potential has two extrema (maximum and minimum) which merge at the point of inflection for $|{\cal L}|=|{\cal L}_{\text{min}}|$ [see Fig.~\ref{f2}, plot (${\bf b}$)]. For $|{\cal L}|<|{\cal L}_{\text{min}}|$ the effective potential has no extrema.
\item[$q=0$] The function $f$ diverges at $x=2$. It has one extremum (minimum) at $x\in(2,+\infty)$, where it is positive. Thus, Eq. \eq{IV.4a} has at most two real roots for appropriate values of $|{\cal L}|>|{\cal L}_{\text{min}}|=2\sqrt{3}$. These roots merge for $|{\cal L}|=|{\cal L}_{\text{min}}|$. Accordingly, the effective potential has two extrema (maximum and minimum) which merge at the point of inflection for $|{\cal L}|=|{\cal L}_{\text{min}}|$ [see Fig.~\ref{f2}, plot (${\bf c}$)]. For $|{\cal L}|<|{\cal L}_{\text{min}}|$ the effective potential has no extrema.
\item[$q\in(0,q_{\text{max}})$] The function $f$ diverges at one point $x=x_{I}$. It has two extrema (maximum and minimum) at $x\in(x_{I},+\infty)$, where it is positive. Thus, Eq. \eq{IV.4a} has at most three real roots for appropriate values of $|{\cal L}|\in(|{\cal L}_{\text{min}}|,|{\cal L}_{\text{max}}|)$, where $|{\cal L}_{\text{min}}|$ and $|{\cal L}_{\text{max}}|$ depend on the value of $q$. Accordingly, the effective potential has three extrema (two maxima and one minimum). The first maximum and the minimum merge at the point of inflection for $|{\cal L}|=|{\cal L}_{\text{min}}|$. The minimum and the second maximum merge at the point of inflection for $|{\cal L}|=|{\cal L}_{\text{max}}|$. This behaviour of the effective potential is illustrated in Fig.~\ref{f2}, plot (${\bf d}$). For $|{\cal L}|<|{\cal L}_{\text{min}}|$ or $|{\cal L}|>|{\cal L}_{\text{max}}|$ the effective potential has only one extremum (maximum).
\item[$q=q_{\text{max}}$] The function $f$ diverges at one point $x=x_{I}$. It has an inflection point at $x\in(x_{I},+\infty)$, where it is positive. One can show that Eq. \eq{IV.4a} has one real root in the region $x\in(x_{I},+\infty)$ for any value of $|{\cal L}|$. Accordingly, the effective potential has only one extremum (maximum). For $|{\cal L}|\approx3.3708$ the effective potential has an inflection point at $x=x_{\text{max}}$. This behaviour of the effective potential is illustrated in Fig.~\ref{f2}, plot (${\bf e}$).
\item[$q>q_{\text{max}}$] The function $f$ has no extrema. It diverges at one point $x=x_{I}$ and monotonically decreases from $+\infty$ to $-\infty$ in the region $x\in(x_{I},+\infty)$. Thus, Eq. \eq{IV.4a} has one real root for any value of $|{\cal L}|$. Accordingly, the effective potential has one extremum (maximum). The behaviour of the effective potential is illustrated in Fig.~\ref{f2}, plot (${\bf f}$).
\end{description}

Thus, we see that the effective potential may allow (depending on the value of $|{\cal L}|$) for stable and marginally stable circular orbits (ISCO's) and finite motion outside the black hole only for $q\in(q_{\text{min}}, q_{\text{max}}]$.

\subsection{Effective potential for null geodesics}

\begin{figure*}[htb]
\begin{center}
\hspace{0cm}
\ba
&&\hspace{0cm}\includegraphics[width=6.0cm]{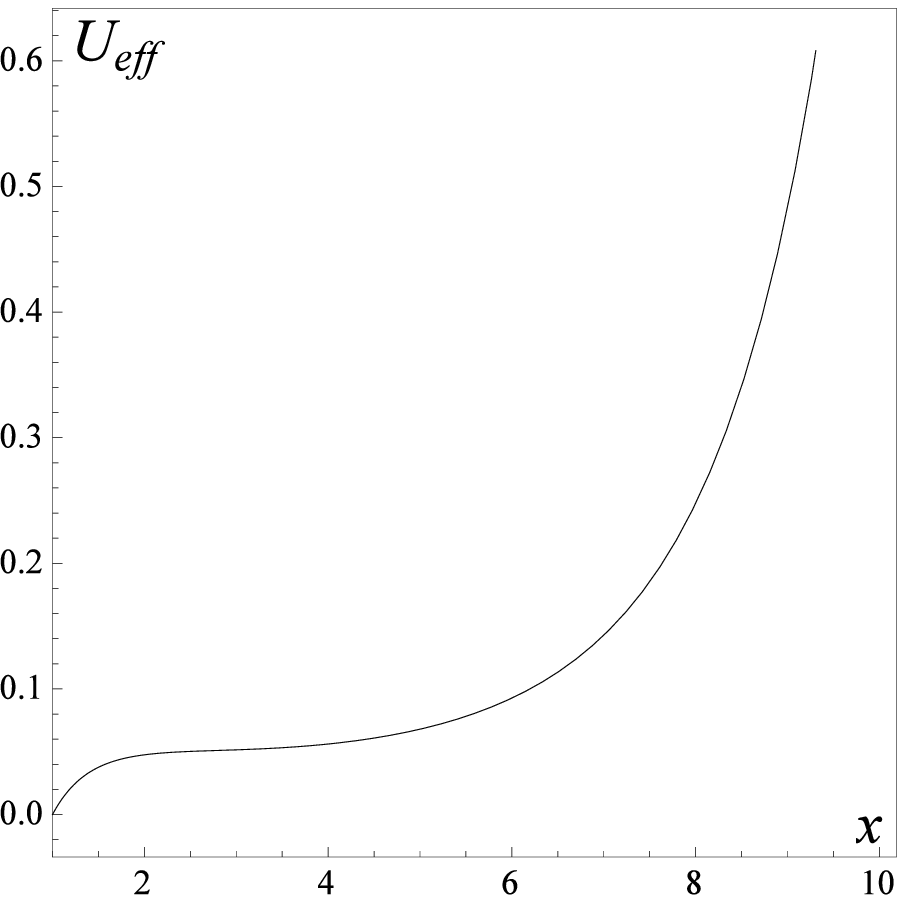}
\hspace{1.5cm}\includegraphics[width=6.0cm]{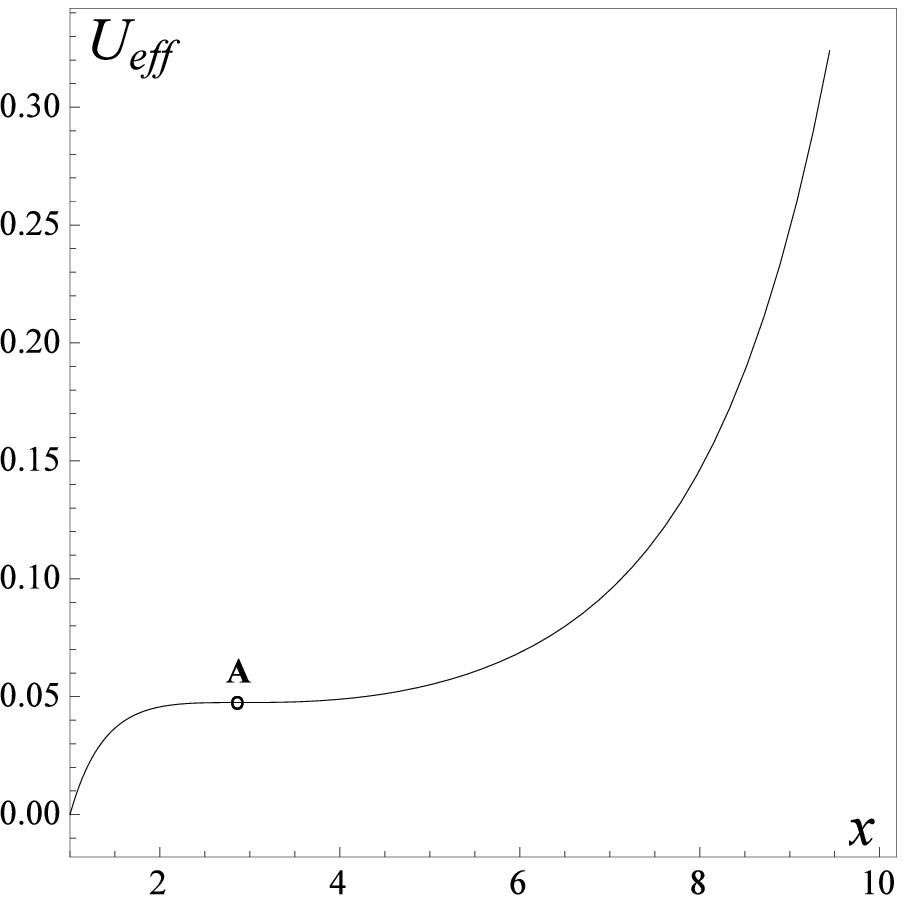}\non\\
&&\hspace{3cm}({\bf a})\hspace{7.1cm}({\bf b})\non\\
&&\hspace{0cm}\includegraphics[width=6.0cm]{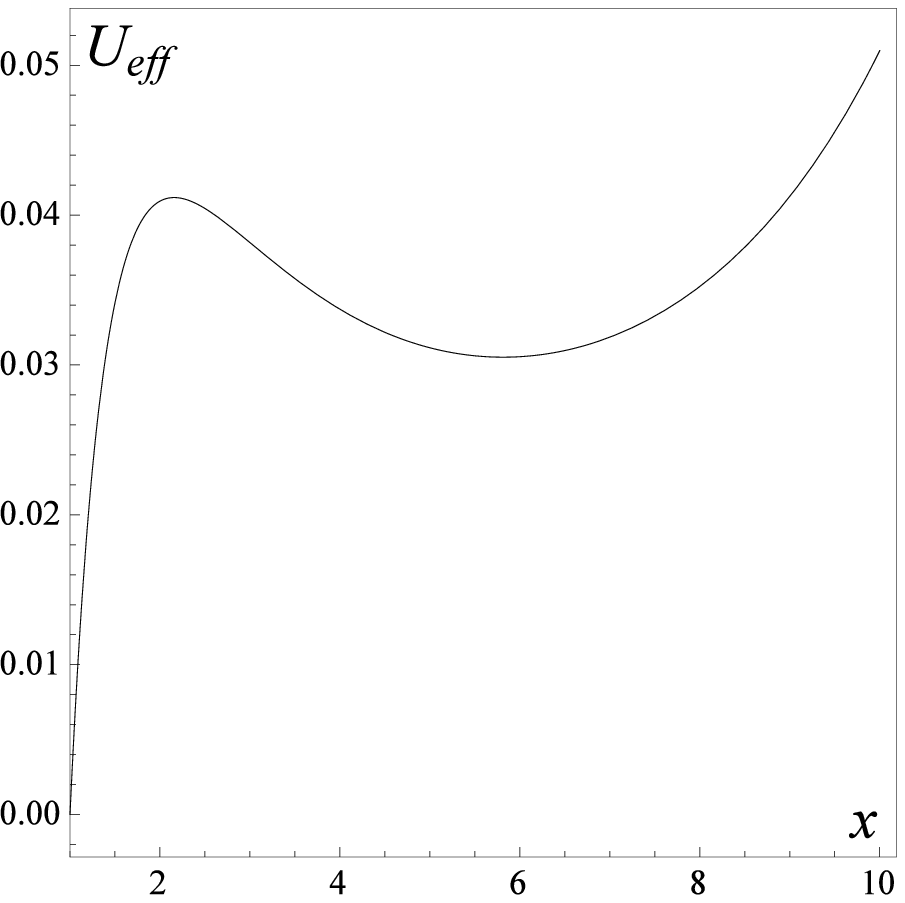}
\hspace{1.5cm}\includegraphics[width=6.0cm]{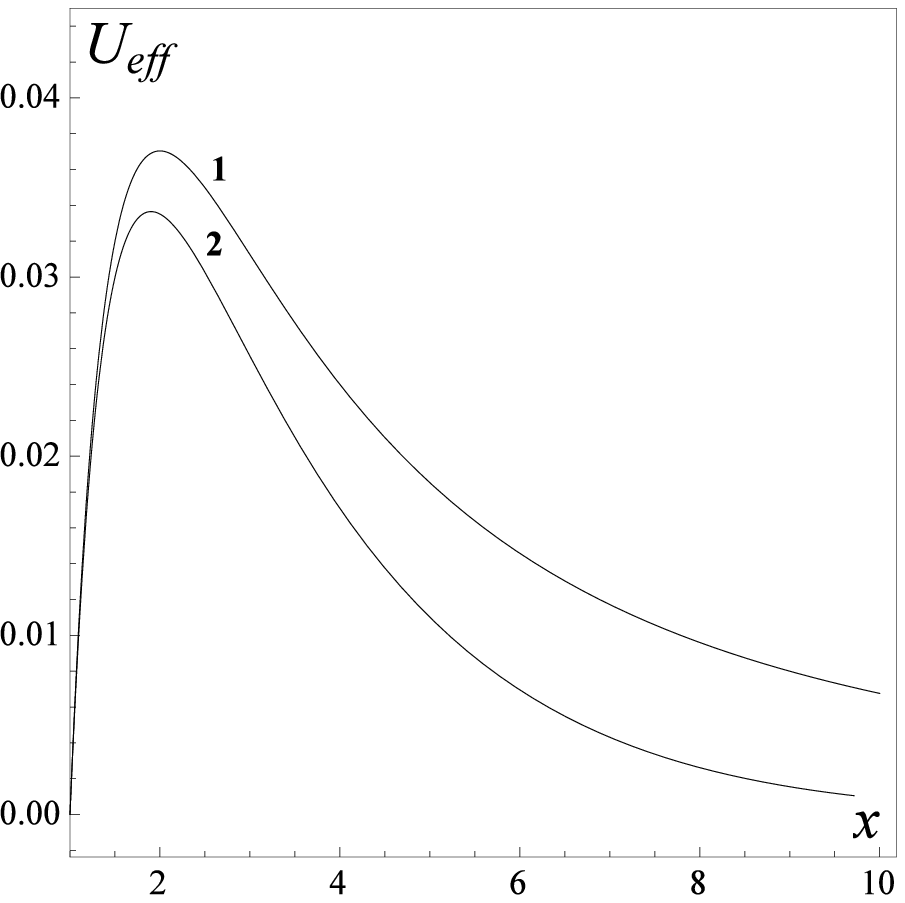}\non\\
&&\hspace{3cm}({\bf c})\hspace{7.1cm}({\bf d})\non\\
\ea
\caption{The effective potential for null geodesics with ${\cal L}=1$. Plot (${\bf a}$): $q<q_{\text{min}}$, $q=-0.0250$. Plot (${\bf b}$): $q=q_{\text{min}}$. The inflection point $A$ is at $x=x_{\text{min}}\approx 2.8794$. Plot (${\bf c}$): $q\in(q_{\text{min}},0)$, $q=-0.01$. Plot (${\bf d}$): $q\geq 0$. 
Curve 1: $q=0$; Curve 2: $q=0.01$.} \label{f3}
\end{center}
\end{figure*}

For null geodesics, the effective potential reads
\be\n{IV.10}
U_{\text{eff}}=\frac{(x-1)}{(x+1)^{3}}{\cal L}^{2}\,e^{4\bar{\cu}}\,.
\ee
Considering again a quadrupole distortion as in \eq{IV.2}, let us now study properties of the effective potential in the region external to the black hole $x\in(1,+\infty)$. We have
\be\n{IV.11}
\lim_{x\to1^{+}}U_{\text{eff}}=0^{+}\hh \lim_{x\to+\infty}U_{\text{eff}}=\begin{cases}
+\infty\hh q<0\\
0^{+}\hh q\geq0\,\,.\\
\end{cases}
\ee
The effective potential is again positive and continuous in this region. In this case, the continuity of $U_{\text{eff}}$ implies that there is an even number of extrema for $q<0$ and an odd number of extrema for $q\geq0$. Again, we seek extrema of $U_{\text{eff}}$. An equation for the extrema $U_{\text{eff},x}=0$ is equivalent to 
\be\n{IV.12}
x-2+2qx(x^{2}-1)=0\,,
\ee
which does not depend on the value of ${\cal L}$. This is precisely the condition where the function $f$ defined in \eq{IV.4b} diverges. It is equivalent to \eq{IV.6} which is illustrated by curve 1 in Fig.~\ref{f1}. From this curve, we can see that the effective potential has extrema if $q \geq q_{\text{min}}$. Using Fig.~\ref{f1} we deduce properties of the effective potential in the region $x\in(1,+\infty)$. They are the following:
\begin{description}
\item[$q<q_{\text{min}}$] No real roots of \eq{IV.12} exist. Thus, the effective potential does not possess any extrema and grows monotonically [see Fig.~\ref{f3}, plot (${\bf a}$)].
\item[$q=q_{\text{min}}$] There is one real root of \eq{IV.12} at $x=x_{\text{min}}$. Thus, the first derivative (as well as the second one) of the effective potential vanishes at one point (point of inflection) when $x=x_{\text{min}}\approx2.8794$ [see Fig.~\ref{f3}, plot (${\bf b}$)].
\item[$q\in(q_{\text{min}},0)$] There are two real roots of \eq{IV.12}. The effective potential has a maximum at $x\in(1,x_{\text{min}})$ and a minimum at $x\in(x_{\text{min}},+\infty)$ [see Fig.~\ref{f3}, plot (${\bf c}$)].
\item[$q=0$] There is one root, $x=2$, of \eq{IV.12}  where the effective potential has a maximum [see Fig.~\ref{f3}, plot (${\bf d}$)].
\item[$q>0$] There is one real root of \eq{IV.12} at $x\in(1,2)$ where the effective potential has a maximum [see Fig.~\ref{f3}, plot (${\bf d}$)].
\end{description}
Thus, we see that the effective potential will allow for an ISCO and bound orbits of null geodesics outside the black hole only for $q\in[q_{\text{min}},0)$.

\section{Equatorial geodesics}
 
Let us now study equatorial geodesics. We shall consider the case of timelike geodesics, $\varepsilon=-1$, and null geodesics, $\varepsilon=0$, separately. 

\subsection{Equatorial timelike geodesics}

One can see an effect of the distortion field on equatorial timelike geodesics on the example of ISCO's. It is well known that in the case of undistorted Schwarzschild black hole $r\ind{ISCO}=6\,m$ which, according to the transformation \eq{II.9}, corresponds to $x\ind{ISCO}=5$. Due to the distortion field, $r\ind{ISCO}$ changes its value. In the case of a quadrupole distortion, the change is defined by both the sign and the value of a quadrupole moment $q$. According to the plots presented in Fig.~\ref{f2}, we see that for $q\in(q_{\text{min}},0)$, $x\ind{ISCO}<5$, while for $q\in(0,q_{\text{max}}]$, $x\ind{ISCO}>5$. There are no ISCOs for $q>q_{\text{max}}$. 

Let us now study the kinematics of a massive particle in the limit $q\to q_{\text{min}}^{+}$. Let us calculate the linear velocity $\hat{v}$ of a massive particle in an equatorial circular orbit as measured by a stationary observer located at the same radius. In the coordinate basis, the 4-velocity of the particle is given by
\be\n{V.13}
u^{\alpha}=\gamma(1,0,0,\Omega)\,,
\ee
where $\gamma$ is the gamma factor and 
\be\n{V.13a}
\Omega=\left.\frac{\dot{\phi}}{\dot{T}}\right|_{y=0}=\frac{\cal L}{\cal E}\frac{(x-1)}{(x+1)^{3}}e^{4\bar{\cu}}
\ee
is the angular velocity of the particle. The tetrad related to the observer reads
\ba\n{V.14}
e^{\hat{T}}_{\alpha}&=&\sqrt{-g\ind{TT}}\,\delta^{T}_{\alpha}\hh e^{\hat{x}}_{\alpha}=\sqrt{g_{xx}}\,\delta^{x}_{\alpha}\,,\non\\
e^{\hat{y}}_{\alpha}&=&\sqrt{g_{yy}}\,\delta^{y}_{\alpha}\hh
e^{\hat{\phi}}_{\alpha}=\sqrt{g_{\phi\phi}}\,\delta^{\phi}_{\alpha}\,.
\ea
In the observer's frame, the particle's 4-velocity is
\be\n{15}
u^{\hat{\alpha}}=e^{\hat{\alpha}}_{\alpha}\,u^{\alpha}=\hat{\gamma}(1,0,0,\hat{v})\,.\,
\ee
From this, we find that the linear velocity of the particle in a circular orbit measured by the stationary observer located at the same radius is
\be\n{V.16}
\hat{v}=\frac{\cal L}{\cal E}\frac{(x-1)^{1/2}}{(x+1)^{3/2}}e^{2\bar{\cu}}\,.
\ee
Using the expression \eq{III.11a} for a circular orbit ($\dot{x}=0$), and the expressions \eq{IV.1} and \eq{IV.4a} we derive
\be\n{V.17}
\hat{v}=\frac{\text{sign}({\cal L})\sqrt{f}}{\sqrt{(x+1)^{2}+f}}\,,
\ee
This expression implies that the observed velocity approaches the speed of light ($\hat{v}\to\pm1$) when $f$ diverges. This corresponds to curve 1 in Fig.~\ref{f1}. Thus bounded timelike geodesics do not exist along this curve which includes $q=q_{min}$.

Let us now find the corresponding values of $x\ind{ISCO}$ defined by the range $q\in(q_{\text{min}}, q_{\text{max}}]$. According to Fig.~\ref{f1} and Fig.~\ref{f2}, plot (${\bf b})$, the minimal value of $x\ind{ISCO}$ corresponds to $q=q_{\text{min}}^{+}$ [cf. Eq. \eq{IV.9a}],
\be\n{V.1}
x\ind{ISCOmin}=\lim_{q_{-}\to q_{\text{min}}^{+}}x(q_{-})=1+2\cos(\pi/9)\approx2.8794\,.
\ee
Using the transformation \eq{II.9}, we find
\be\n{V.2}
r\ind{ISCOmin}=2m[1+\cos(\pi/9)]\approx3.8794\,m\,.
\ee
This value is approximately $29\%$ larger than the radius of the photon sphere of the Schwarzschild black hole, $r\ind{photon}=3\,m$. 

In order to see the effect of distortion in an invariant way, let us calculate the proper distance from the ISCO to the black hole horizon $\ell_{\text{min}}$ and the ISCOs circumference ${\cal C}_{\text{min}}$. We consider adiabatic distortion, so that the distorted black hole horizon area 
\be\n{V.3}
A_h=16\pi m^2 e^{-2u_0}
\ee
is constant, which is equal to the horizon surface area of an undistorted Schwarzschild black hole. We define the proper distance from the ISCO to the black hole horizon and its circumference in units of the radius $r_h$ corresponding to the area $A_h$,
\be\n{V.4}
r_h=\left(\frac{A_h}{4\pi}\right)^{1/2}=2me^{-u_0}\,.
\ee
Using the metric \eq{II.1}, where we put $t=const$, $y=0$, and $\phi=const$, and the expressions \eq{II.4d}, we derive the proper distance $\ell_{\text{min}}$ from the black hole horizon ($x=1$) to the ISCO ($x=x\ind{ISCOmin}$),
\ba
\ell_{\text{min}}&=&\frac{1}{2}\int_1^{x\ind{ISCOmin}}dx\sqrt{\frac{x+1}{x-1}}e^{\frac{q_{\text{min}}}{2}(x^2-4x+1)+\frac{q_{\text{min}}^2}{4}(x^2-1)^2}\non\\
&\approx& 2.2722\,,\n{V.5}
\ea
where we used the expressions \eq{III.9a} and \eq{III.9b} for a quadrupole distortion. In the absence of distortion we derive
\be\n{V.6}
\ell_{\ind{Sch}}=\frac{1}{2}\int_1^5 dx\sqrt{\frac{x+1}{x-1}}=\sqrt {6}+\frac{1}{2}\ln  \left( 5+2\,\sqrt {6} \right)\approx 3.5957\,.
\ee
The proper distance form the distorted black hole's horizon to the corresponding ISCO is $37\%$ less than that of an undistorted Schwarzschild black hole. 

Let us now calculate the circumference ${\cal C}_{\text{min}}$ of the ISCO. Using the metric \eq{II.1}, where we put $t=const$, $y=0$, and $x=x\ind{ISCOmin}$, and the expression \eq{III.9a}, we derive
\be\n{V.7} 
{\cal C}_{\text{min}}=\pi(x\ind{ISCOmin}+1)e^{\frac{q_\text{min}}{2}(x\ind{ISCOmin}^2+1)}\approx 11.0575\,.
\ee 
In the absence of distortion one has
\be\n{V.8}
{\cal C}_{\ind{Sch}}=6\pi\approx 18.8496\,.
\ee
The circumference of the distorted black hole's ISCO is $41\%$ less than that of an undistorted Schwarzschild black hole. 

As a summary we conclude that due to the negative quadrupole distortion of the minimal value $q=q_\text{min}^{+}$, the distorted black hole's ISCO is closer to the horizon and its circumference becomes less than that of an undistorted Schwarzschild black hole.  

Let us now consider the case of positive quadrupole moment $q\in(0,q_{\text{min}}]$. According to Fig.~\ref{f1} and Fig.~\ref{f2}, plots (${\bf d})$ and (${\bf e})$, the maximal value of $x\ind{ISCO}$ can be arbitrary large within the validity of the local black hole model (see discussion at the end of Conclusion). Indeed, one can infer that for $q\to 0^{+}$ and an appropriate value of $|{\cal L}|$ the point of inflection corresponding to a merger between the minimum and the second maximum of the effective potential can be arbitrary far from the black hole horizon. On the other side, the $x$-coordinate of the point of inflection corresponding to a merger between the first maximum and the minimum of the effective potential approaches the value of $x=5$. In what follows, we shall consider the point of inflection corresponding to the maximal value of $q=q_{\text{max}}$ and define its $x$-coordinate as $x\ind{ISCOmax}$. Using the results of the previous section and the transformation \eq{II.9}, we find
\be\n{V.9}
x\ind{ISCOmax}\approx6.5018\,,
\ee
and
\be\n{V.10}
r\ind{ISCOmax}\approx7.5018\,m\,.
\ee
This value is approximately $25\%$ larger than the radius of the ISCO of the Schwarzschild black hole, $r\ind{ISCO}=6m$. 

Let us now calculate the proper distance $\ell_{\text{max}}$ from the black hole horizon ($x=1$) to the ISCO ($x=x\ind{ISCOmax}$) and the ISCO circumference ${\cal C}_{\text{max}}$. Replacing $q_{\text{min}}$ with $q_{\text{max}}$ and $x\ind{ISCOmin}$ with $x\ind{ISCOmax}$ in the expressions \eq{V.5} and \eq{V.7} we derive
\be\n{V.11}
\ell_{\text{max}}\approx4.4927\,,
\ee
and
\be\n{V.12}
{\cal C}_{\text{max}}\approx23.7062\,.
\ee
The proper distance from the distorted black hole's horizon to the corresponding ISCO is $25\%$ greater and the ISCO circumference is $26\%$ greater than those of an undistorted Schwarzschild black hole. 

\subsection{Equatorial null geodesics}

We will now examine the effect of the distortion field on equatorial null geodesics. In the case of an undistorted Schwarzschild black hole, no stable circular orbits exist. Circular orbits of radius $r=3\,m$ corresponding to $x=2$ by transformation \eq{II.9}, are possible, but are unstable. However, in the presence of a quadrupole distortion, stable circular orbits exist and the radius of these orbits are defined by the value of the quadrupole moment $q$. According to the plots presented in Fig.~\ref{f3}, we see that for $q\in[q_{\text{min}},0)$, bounded orbits exist. 

According to Fig.~\ref{f1} and Fig.~\ref{f3}, plots (${\bf b})$ and (${\bf c})$, the value of $x\ind{ISCO}$ corresponds to $q=q_{\text{min}}$ which can be found from Eq. \eq{IV.6}. The result is 
\be\n{V.21}
x\ind{ISCO}=x_{\text{min}}=1+2\cos(\pi/9)\approx2.8794\,.
\ee
Using the transformation \eq{II.9}, we find
\be\n{V.22}
r\ind{ISCO}=2m[1+\cos(\pi/9)]\approx3.8794\,m\,.
\ee

As was the case for timelike geodesics, the minimum proper distance from the horizon to the ISCO $\ell_{\text{min}}$ is given by \eq{V.5} and the minimum circumference ${\cal C}_{\text{min}}$ of the ISCO is given by \eq{V.7}.

Due to the negative quadrupole distortion of the minimal value $q=q_\text{min}$, the ISCO of null geodesics is the limiting orbit for ISCO's of a massive test particle approaching the speed of light (see the previous subsection). 

\begin{figure}[htb]
\begin{center}
\hspace{0cm}
\includegraphics[width=7.0cm]{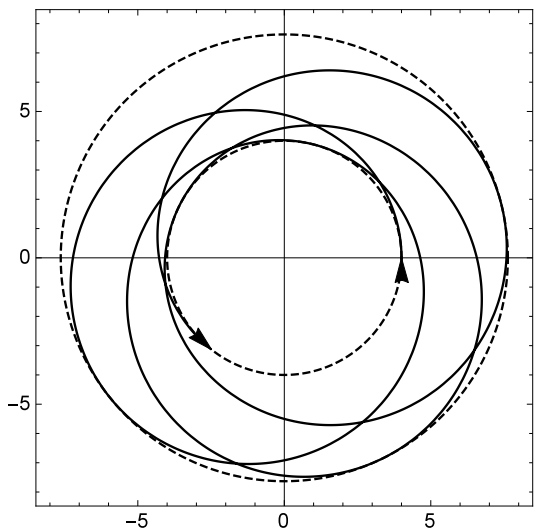}
\caption{Bounded photon orbit with a periastron shift for $q=-0.01$ and ${\cal L}=1$. The orbit's parameters are  $x_{\text{min}}=4$, $x_{\text{max}}\approx7.6308$, and $\lambda\in[0,700]$. The arrows illustrate the initial and final points of the orbit.} \label{f4}
\end{center}
\end{figure}

A bounded photon orbit corresponding to the effective potential in Fig.~\ref{f3}, plot (${\bf c})$ is shown in Fig.~\ref{f4}.

In the case of a positive quadrupole moment $q>0$, no stable circular orbits exist. However, according to Fig.~\ref{f1} and Fig.~\ref{f3}, plot (${\bf d})$, there exists an unstable circular orbit approaching $x=1$ for arbitrarily large $q$. This corresponds to the horizon of the distorted black hole $r=2\,m$.
 
\section{Conclusion}

In this paper we studied timelike and null geodesics in the vicinity of a local static distorted black hole. We restricted ourselves to geodesics lying in the equatorial plane. Such geodesics exist if there are no odd interior multipole moments in the distortion field. This corresponds to a space-time having reflective symmetry across the equatorial plane. We considered a quadrupole distortion defined by the quadrupole moment $q$. 

As a result of our study, we found that there are ISCOs and bound orbits of timelike geodesics for $q\in(q_{\text{min}},q_{\text{max}}]$, where $q_{\text{min}}\approx -0.0210$ and $q_{\text{max}}\approx 2.7086\times10^{-4}$. For $q\in(q_{\text{min}},0)$ the corresponding ISCOs are located closer to the black hole horizon than the ISCO of a Schwarzschild black hole of the mass $m$, which is at $r=6\,m$. The closest ISCO is defined by $q\to q_{\text{min}}^-$ which is at $r\ind{ISCOmin}\approx3.8794\,m.$ This value is approximately $29\%$ larger than the radius of the photon sphere of the Schwarzschild black hole, $r\ind{photon}=3\,m$. We also calculated the proper distance to the distorted black hole horizon and the circumference for the closest ISCO and found that the proper distance is $37\%$ less and the circumference is $41\%$ less than those corresponding to the ISCO of the Schwarzschild black hole. 

For positive values of the quadrupole moment $q\in(0,q_{\text{max}}]$ the effective potential has three extrema, two maxima and one minimum. Thus one class of ISCOs corresponds to a merger between the minimum and the second maximum of the effective potential. As a result, such ISCOs can be located arbitrarily far from the black hole horizon within the region of validity of our solution (see the last paragraph of this section). Another class of ISCOs corresponds to a merger between the first maximum and the minimum of the effective potential. As a result, such ISCOs are located close to a distorted black hole horizon. We calculated that for $q=q_{\text{max}}$ there is the farthermost ISCO of the second class located at $r\ind{ISCOmax}\approx7.5018\,m$. This value is approximately $25\%$ larger than the radius of the ISCO of a Schwarzschild black hole of the mass $m$, $r\ind{ISCO}=6\,m$. We calculated the proper distance to the distorted black hole horizon and the circumference for this ISCO and found that the proper distance is $25\%$ greater and the circumference is $26\%$ greater than those corresponding to the ISCO of the Schwarzschild black hole.

In addition to the ISCOs and bound orbits there are also (unstable) ``static points'' where a massive particle can stay at rest with respect to the distorted black hole. These points are defined by curve 2 in Fig.~\ref{f1}. Note that there is only one point for a given value of $q$. Due to the axial symmetry, this point corresponds to a ``static ring'' around a distorted Schwarzschild black hole.

We found that for $q\in[q_{\text{min}},0)$ there is a null ISCO as well as bound null geodesic orbits  lying in the equatorial plane of a distorted Schwarzshild black hole. The ISCO exists for $q=q_{\text{min}}$ and it is located at $r\ind{ISCO}\approx3.8794\,m$. The proper distance from this ISCO to the distorted black hole horizon is $37\%$ less and its circumference is $41\%$ less than these of ISCO of a timelike geodesics of a Schwarzschild black hole of mass $m$. We illustrated the existence of bound orbits of null geodesics lying in the equatorial plane of the distorted black hole in Fig.~\ref{f4}. These orbits have periastron shift. They create a ``corridor of light'' around a distorted black hole, in its equatorial plane.

The existence of the ISCO and bound orbits of null geodesics, as well as ISCOs of timelike geodesics which are located closer to the distorted black hole horizon than the ISCO of a Schwarzschild black hole, can be intuitively explained by Newtonian picture. In Newtonian gravity a gravitational potential whose multipole expansion is dominated by an interior quadrupole moment $q_{N}$ can be modelled by two equal pointlike masses $\mu$ located on the $z$-axis at the distance $d$ from the origin and an infinitesimally thin ring of the mass $M$ and radius $R$ located at the plane perpendicular to the $z$-axis and centred at the origin. In this case we have $q_{N}=M/(2R^3)-2\mu/d^3$. Thus, if the contribution of the point-like masses to the gravitational field is greater than that of the ring, then $q_N<0$, otherwise, $q_N\geq0$. If $q_N<0$, then there is a net force acting on a particle and directed towards the $z$-axis. This force creates the potential barrier at some distance from the black hole horizon, as in the plots (${\bf a}$), and (${\bf b}$) in Fig.~\ref{f2} and (${\bf a}$), (${\bf b}$), (${\bf c}$) in Fig.~\ref{f3}. The other potential barrier, which is closer to the black hole horizon, is due to the angular motion. In a similar way, one can explain the existence of the ``static ring'' for $q\in(0,q_{\text{max}}]$. In this case $q_{N}>0$ and the contribution of the ring to the gravitational potential is greater than that of the pointlike masses. As a result, there is a net force directed to the ring, outward from the black hole. This force balances the black hole's gravitational pull at the ``static ring'' around the black hole.  

In order to justify the distortion in this way, its contribution to the space-time curvature in the vicinity of the distorted black hole horizon should be small compared to that of the black hole. To estimate a contribution of the distortion to the space-time curvature, let us consider the quadrupole distortion at the black hole horizon, on the equatorial plane. On a static black hole horizon the Kretschmann scalar curvature invariant  is ${\cal K}|_{\ind{Horizon}}=12K^2$, where $K$ is the Gaussian curvature of the horizon two-dimensional spacelike surface (for details see, e.g., \cite{FrSan,FS,AFS,Abdolrahimi,Shoom}). For the metric \eq{II.5}, $K=(1+2q)\exp(2q)$, and for $|q|\ll1$ one has $K\approx1+4q+6q^2$. This suggests that for $q\in[q_{\text{min}}, q_{\text{max}}]$ our interpretation  is justified. Such a distortion slightly modifies the space-time geometry in the vicinity of the black hole's horizon and results in the deformation of the timelike and null geodesics.

Note that the presence of an extra spatial dimension has a similar effect on null geodesics around a uniform black string \cite{Gonzalez}. As it was shown in \cite{Gonzalez}, due to the extra fifth (compact) spatial dimension, a photon acquires a positive test mass. As a result, the corresponding effective potential is exactly the same as that for a massive test particle moving around a Schwarzschild black hole in its equatorial plane. However, the photon's mass is proportional to the speed of propagation of the photon in the fifth dimension. Thus, bound orbits of null geodesics are not confined to an equatorial plane of the Schwarzschild black hole.

In this paper we have restricted our attention to geodesics in the equatorial plane. However it would be interesting to study general timelike and null geodesics around  a distorted Schwarzschild black hole as well as those around a distorted Kerr black hole.  
 
\appendix*

\section{Einstein equations}

In this Appendix we present the vacuum Einstein equations $R_{\mu\nu}=0$ for the metric \eq{II.1}. These equations reduce to the following equations for the distortion fields $U(x,y)$ and $V(x,y)$ in the prolate spheroidal coordinates:
\ba
&&\hspace{-1.0cm}(1-x^{2})U_{,xx}-2xU_{,x}-(1-y^{2})U_{,yy}+2yU_{,y}=0\,,\n{A1}\\
V_{,x}&=&\frac{(1-y^{2})}{(x^{2}-y^{2})}\left(x\left[(x^{2}-1)U_{,x}^{2}-(1-y^{2})U_{,y}^{2}\right]\right.\non\\
&-&\left.2y(x^{2}-1)U_{,x}U_{,y}+2xU_{,x}-2yU_{,y}\right)\,,\n{A2}\\
V_{,y}&=&\frac{1}{(x^{2}-y^{2})}\left(y(x^{2}-1)\left[(x^{2}-1)U_{,x}^{2}-(1-y^{2})U_{,y}^{2}\right]\right.\non\\
&+&2x(x^{2}-1)(1-y^{2})U_{,x}U_{,y}\n{A3}\\
&+&\left.2y(x^{2}-1)U_{,x}+2x(1-y^{2})U_{,y}\right)\,.\non 
\ea  
The functions \eq{II.2a}--\eq{II.3} solve these equations (see, e.g., \cite{Man,ManQ}).

\begin{acknowledgments}

This research was supported  by the Natural Sciences and Engineering
Research Council of Canada Discovery Grant 261429-2013.

\end{acknowledgments}


\begin{thebibliography}{99}

\bibitem{Carter} B. Carter, in {\em Black Holes: Les Houches 1972}, edited by C. DeWitt and B. S. DeWitt (Gordon and Breach Science Publishers, Inc., New York, 1973).
\bibitem{DK} D.~Kubiznak, arXiv:0809.2452 [gr-qc].
\bibitem{FrK} V.~P.~Frolov and D.~Kubiznak, Class.\ Quant.\ Grav.\  {\bf 25},154005 (2008).
\bibitem{HE} S.~W.~ Hawking and G.~F.~R.~Ellis, {\em The large scale structure of space-time}, (Cambridge University Press, New York, NY, 1973). 
\bibitem{Tipler1} F.~J.~Tipler, Phys.\ Rev.\ D {\bf 9}, 2203 (1974).
\bibitem{Tipler2} F.~J.~Tipler, Phys.\ Rev.\ Lett.\  {\bf 37} (1976) 879.
\bibitem{Chandrabook} S. Chandrasekhar, {\em The Mathematical Theory of Black Holes}, (Clarendon Press, Oxford, 1983).
\bibitem{Hackmann1} E.~Hackmann, V.~Kagramanova, J.~Kunz and C.~Lammerzahl, Phys.\ Rev.\ D {\bf 78}, 124018 (2008) [Phys.\ Rev.\  {\bf 79}, 029901 (2009)].
\bibitem{Kagramanova} V.~Kagramanova, J.~Kunz, E.~Hackmann and C.~Lammerzahl, Phys.\ Rev.\ D {\bf 81}, 124044 (2010).
\bibitem{Diemer} V.~Diemer and J.~Kunz, Phys.\ Rev.\ D {\bf 89}, 084001 (2014).
\bibitem{Hackmann2} E.~Hackmann, C.~Lammerzahl, V.~Kagramanova and J.~Kunz, Phys.\ Rev.\ D {\bf 81}, 044020 (2010).
\bibitem{Perlick} V.~Perlick, arXiv:1010.3416 [gr-qc].
\bibitem{FZ} V.~P.~Frolov and A.~Zelnikov, {\em Introduction to Black Hole Physics}, (Oxford University Press, 2011).
\bibitem{Enolskii} V.~Enolskii, B.~Hartmann, V.~Kagramanova, J.~Kunz, C.~Laemmerzahl and P.~Sirimachan, Phys.\ Rev.\ D {\bf 84}, 084011 (2011).
\bibitem{Shohreh1} S.~Abdolrahimi, R.~B.~Mann and C.~Tzounis, Phys.\ Rev.\ D {\bf 91}, 084052 (2015).
\bibitem{Shohreh2} S.~Abdolrahimi, R.~B.~Mann and C.~Tzounis, arXiv:1510.03530 [gr-qc].  
\bibitem{Garcia} A.~Garc'a, E.~Hackmann, J.~Kunz, C.~LŠmmerzahl and A.~Mac'as, J.\ Math.\ Phys.\  {\bf 56}, 032501 (2015).
\bibitem{Mashhoon1} B.~Mashhoon, F.~W.~Hehl and D.~S.~Theiss, Gen.\ Rel.\ Grav.\  {\bf 16}, 711 (1984).
\bibitem{Mashhoon2} B. Mashhoon, Phys. Rev. D {\bf 10}, 1059 (1974).
\bibitem{Mashhoon3} B. Mashhoon, Nature {\bf 250}, No. 5464, 316 (1974).
\bibitem{FrSh1} V.~P.~Frolov and A.~A.~Shoom, Phys.\ Rev.\ D {\bf 84}, 044026 (2011).
\bibitem{FrSh2} V.~P.~Frolov and A.~A.~Shoom, Phys.\ Rev.\ D {\bf 86}, 024010 (2012).
\bibitem{FN} V. P. Frolov and I. D. Novikov, {\em Black Hole Physics: Basic Concepts and Recent Developments} (Kluwer Academic Publishers, Dodrecht-Boston-London, 1998).
\bibitem{MTW} C. W. Misner, K. S. Thorne and J. A. Wheeler,
{\em Gravitation},   W. H. Freeman and Co., San Francisco, (1973).
\bibitem{Teo} E. Teo, Gen. Relativ. Gravit. {\bf 35}, 1909 (2003).
\bibitem{Cederbaum1} C.~Cederbaum, arXiv:1406.5475 [math.DG].
\bibitem{Cederbaum2} C.~Cederbaum and G.~J.~Galloway, arXiv:1504.05804 [math.DG].
\bibitem{Cederbaum3} C.~Cederbaum and G.~J.~Galloway, arXiv:1508.00355 [math.DG].
\bibitem{PV1} J. Podolsk\'y and K. Vesel\'y, Class. Quantum Grav. {\bf 15}, 3505 (1998).
\bibitem{PV2} J. Podolsk\'y and K. Vesel\'y, Phys. Lett. A {\bf 271}, 368 (2000).
\bibitem{Konoplya} R.~A.~Konoplya and Y.~C.~Liu, Phys.\ Rev.\ D {\bf 86}, 084007 (2012). 
\bibitem{FS} V. P. Frolov and A. A. Shoom, Phys. Rev. D {\bf 76}, 064037 (2007).
\bibitem{AFS} S. Abdolrahimi, V. P. Frolov, and A. A. Shoom, Phys. Rev. D {\bf 80}, 024011 (2009).
\bibitem{AAS} A. A. Shoom, Phys. Rev. D {\bf 91}, 064030 (2015).
\bibitem{Man} N. Breton, T. E. Denisova, V. S. Manko, Phys. Lett. A {\bf 230}, 7 (1997).
\bibitem{ManQ} N. Breton, A. A. Garcia, V. S. Manko, and T. E. Denisova, Phys. Rev. D {\bf 57}, 3382 (1998).
\bibitem{GH} R. Geroch and J. B. Hartle, J. Math. Phys. {\bf 23}, 680 (1982).
\bibitem{Gonzalez} P.~A.~Gonzalez, M.~Olivares and Y.~Vasquez, arXiv:1511.08048 [gr-qc].
\bibitem{SuenI} W. Suen, Phys. Rev. D {\bf 34}, 3633 (1986).
\bibitem{Thorne} K. S. Thorne, Rev. Mod. Phys. {\bf 52}, 299 (1980).
\bibitem{Ger1} R. Geroch, J. Math. Phys. {\bf 11}, 1955 (1970).
\bibitem{Ger2} R. Geroch, J. Math. Phys. {\bf 11}, 2580 (1970).
\bibitem{Han} R. O. Hansen, J. Math. Phys. {\bf 15}, 46 (1974).
\bibitem{Que} H. Quevedo, Phys. Rev. D {\bf 39}, 2904 (1989).
\bibitem{SuenII} W. Suen, Phys. Rev. D {\bf 34}, 3617 (1986). 
\bibitem{Beig} R. Beig and W. Simon, J. Math. Phys. {\bf 24}, 1163 (1983).
\bibitem{Gursel} Y. G\"ursel, Gen. Relativ. Gravit. {\bf 15}, 737 (1983).
\bibitem{Eh} J. Ehlers, in {\em Grundlagenprobleme der modernen Physik}, edited by J. Nitsch {\em et al.} (BI Verlag, Mannheim, 1981), p. 65.
\bibitem{Qu} H. Quevedo, Phys. Rev. {\bf 39}, 2904 (1989).
\bibitem{Dor}  A. G. Doroshkevich, Ya. B. ZelÕdovich and I. D. Novikov, Zh. Eksp. Teor. Fiz. {\bf 49}, 170 (1965).
\bibitem{Israel} W. Israel, Phys. Rev. {\bf 164}, 1776 (1967).
\bibitem{FrSan} V. P. Frolov and N. Sanchez, Phys. Rev. D {\bf 33}, 1604 (1986).
\bibitem{Abdolrahimi} S.~Abdolrahimi and A.~A.~Shoom, Phys.\ Rev.\ D {\bf 83}, 104023 (2011).
\bibitem{Shoom} A.~A.~Shoom, Phys.\ Rev.\ D {\bf 91}, 024019 (2015).

\end{thebibliography}
\end{document}